\documentclass[a4paper]{jpconf}
\usepackage{graphicx}

\begin{document}
\title{Weak antilocalization in (111) thin films of a topological crystalline insulator SnTe}

\author{R Akiyama, K Fujisawa, R Sakurai, and S Kuroda}

\address{Institute of Materials Science, University of Tsukuba, 1-1-1 Tennoudai, Tsukuba, Ibaraki 305-8573, Japan}

\ead{akiyama@ims.tsukuba.ac.jp}

\begin{abstract}

We grew single-crystal thin films of a topological crystalline insulator (TCI) SnTe with a smooth surface  at the atomic scale by molecular beam epitaxy (MBE).
In the magnetoresistance (MR) measurement, we observed both positive and negative components near zero magnetic field at lowest temperatures of 2 - 3 K, while we observed only a negative MR at elevated temperatures of 6 - 10 K. The positive MR is attributed to the weak antilocalization (WAL) in the transport through the topological surface state (SS), demonstrating $\pi$ berry phase which is essential to the topological SS, while the negative MR to the weak localization (WL) in the transport through the bulk state (two-dimensional bulk subbbands). The absolute value of the prefactor $ \alpha $ deduced from the fitting of the observed positive MR to the Hikami-Larkin-Nagaoka equation was much smaller than expected from the number of transport channel of the SS, suggesting the coupling of the SS to the bulk state.
\end{abstract}

\section{Introduction}
Topological insulator (TI) is a new class of materials. The surface state (SS) of TIs has a band structure of Dirac-cone shape where the spin of carriers is locked with their momentum except for $k$=0\cite{Fu}. The SS of TIs is metallic protected by time-reversal symmetry and exceedingly high mobility is expected in the transport of mass-less Dirac fermions in the SS. The topological SS has so far been investigated intensively, for example, in Bi$_{1-x}$Sb$_{x}$\cite{Hsieh1}, Bi$_{2}$Se$_{3}$\cite{Moore}, Bi$_{2}$Te$_{3}$\cite{Chen}, and TlBiSe$_{2}$\cite{Kuroda} through electrical transport measurements and the angle-resolved photoemission spectroscopy (ARPES). Recent explorations of TIs have opened the door of a new type of TIs, that is, topological crystalline insulators (TCIs), in which the topological SS is protected by the mirror symmetry (reflection symmetry) of the crystal\cite{Hsieh3}\cite{Fu2}, instead of the time-reversal symmetry.
SnTe was predicted to be one of the TCIs whose SS is protected by mirror symmetry with respect to the \{111\}, \{001\}, and \{110\} planes of the rock-salt crystal structure. The APRES measurement
confirmed the topological phase in the (001) and (111) surfaces of SnTe by observing Dirac-cone band dispersion derived from the SS\cite{Tanaka}\cite{Tanaka2} as well as Pb$_{1-x}$Sn$_{x}$Te($x$ $<$ 0.25)\cite{Xiu} and in the (001) surface of Pb$_{0.73}$Sn$_{0.27}$Se\cite{Dziawa}. In addition, the electrical transport properties which are characteristics of two-dimensional (2D) SS have been reported, such as the Shubnikov-de Haas (SdH) oscillation\cite{Taskin}, and the Altshuler-Aronov-Spivak (AAS)/ the Aharonov-Bohm (AB) effect\cite{Safdar}.

In the present study, we investigated the magneto-transport properties in a SnTe(111) thin film, which was grown by molecular beam epitaxy (MBE). In the transverse magnetoresistance (MR), we observed a positive MR, which is attributed to the weak antilocalization (WAL) in the transport through the SS. The WAL in the SS of TIs is considered to arise from a different mechanism from the case of the conventional transport; the prohibition of backscattering due to the accumulated $\pi$ Berry phase of Dirac fermions is essential for the WAL instead of the spin-orbit interaction (SOI)\cite{Nomura}. Therefore, the investigation of the WAL in TIs or TCIs is useful for gaining an insight into the transport properties of the 2D SS\cite{He}\cite{Ning}.
In the case of SnTe, the SS has not been confirmed sufficiently from the viewpoint of the electrical transport because of the recent recognition as a TCI.
In this article, we report the observation of a positive MR originating from the WAL in the SS in a (111) thin film of single-crystal SnTe. This result suggests that the surface transport of 2D SS of SnTe can be detected even in the existence of a large contribution to the electrical transport of high-density bulk carriers due to Sn vacancies.

\section{Experimental}
\begin{figure}[h]
\begin{center}
\begin{minipage}{17pc}
\begin{center}
\includegraphics[width=18pc]{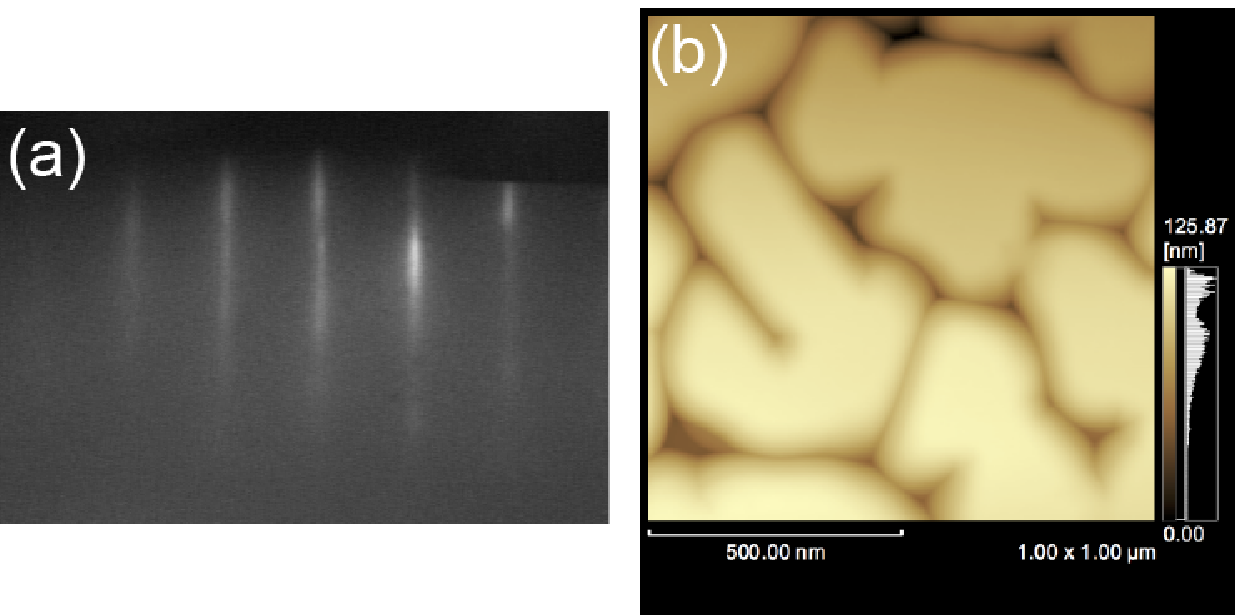}
\caption{\label{Fig1}(a) The RHEED image along [1-10] direction of the grown SnTe film. (b) The atomic force microscope (AFM) image of the surface of the grown SnTe film.}
\end{center}
\end{minipage}\hspace{2pc}%
\begin{minipage}{17pc}
\begin{center}
\includegraphics[width=14pc]{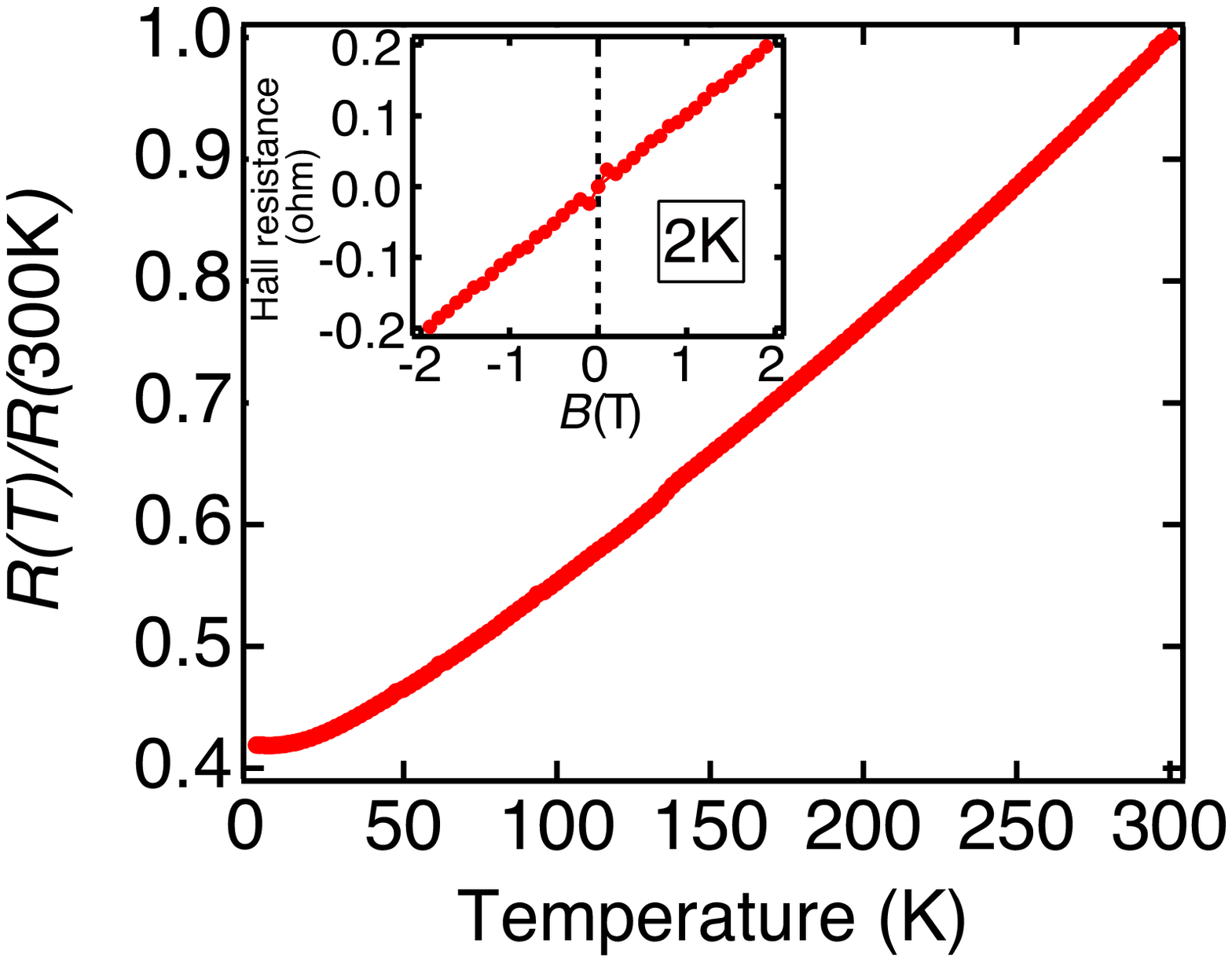}
\caption{\label{Fig2}The temperature dependence of the resistance of the SnTe film. Inset: The magnetic field dependence of the Hall resistance measured in the Van der Pauw configuration at 2K.}
\end{center}
\end{minipage} 
\end{center}
\end{figure}

Thin films of the (111) plane of single-crystal SnTe were epitaxially grown on BaF$_{2}$(111) substrates by molecular beam epitaxy (MBE) equipped with elementary sources of Sn and Te. The substrate temperature was kept at 220$^\circ$C during the growth of SnTe.
The molecular beam flux ratio of Sn/Te was 1/20 (Te-rich condition). The thickness of the grown SnTe film was 115 nm. 
Figure \ref{Fig1} (a) and (b) show a reflection high energy electron diffraction (RHEED) image along [1-10] direction and an atomic force microscope (AFM) image of the surface of the grown SnTe film, respectively. As shown in these images, the surface of the SnTe film is smooth in the range of the atomic scale although the macroscopic morphology shows a characteristic structure. The surface morphology was found to be quite sensitive to the molecular beam flux ratio Sn/Te and the substrate temperature during growth. The electrical transport measurements under magnetic fields  were performed in the standard four-probe and the Van der Pauw configurations using Quantum Design Physical Properties Measurement System (PPMS).

\section{Results and discussion}
\begin{figure}[h]
\begin{center}
\begin{minipage}{35pc}
\begin{center}
\includegraphics[width=29pc]{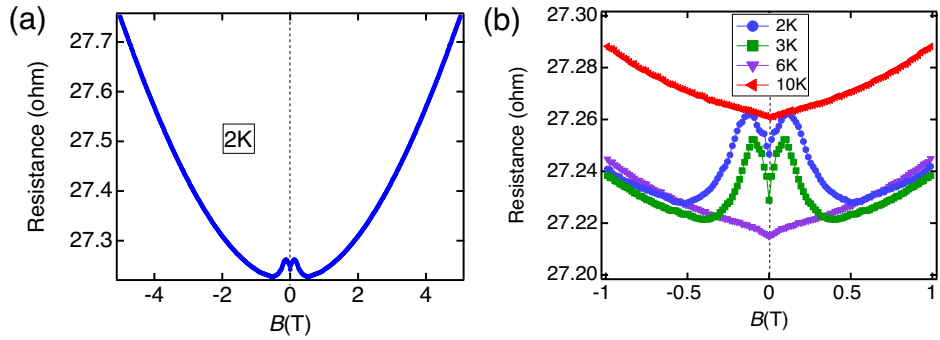}
\caption{\label{Fig3}(a)The magnetic field dependence of the longitudinal resistance measured in the four-probe configuration at 2 K. The component of the odd function of magnetic field was subtracted from the raw data. (b)The magnetoresistance (MR) in the range of $\left| B \right| \leq$ 1 T at 2 - 10 K.}
\end{center}
\end{minipage} 
\end{center}
\end{figure}

Figure \ref{Fig2} shows the temperature dependence of the resistance of the grown SnTe film. 
In the figure, the value of resistance is normalized by the value at 300 K. The temperature dependence shows a metallic behavior with decrease in resistance as the temperature decreases. The inset of Fig. \ref{Fig2} shows the Hall resistance, which was obtained by extracting a component of the odd function of magnetic field from the raw data. From the Hall coefficient, it was deduced that the carrier type is a hole and the hole density is 5.0$\times$10$^{20}$cm$^{-3}$. This value is approximately consistent with the previous report\cite{Tanaka}.
Figure \ref{Fig3} shows the magnetic field dependence of the longitudinal resistance at 2 K under a magnetic field perpendicular to the film plane (transverse MR). In the MR curve, a component of the odd function of magnetic field, which originates from the Hall voltage due to a small deviation of the alignment of the electrodes, was subtracted. As shown in the figure, the MR curve depends on the magnetic field $B$ in a quadratic way in the magnetic field range of $\left| B \right|$ $\geq \sim$ 2.5 T, and there appears a small peak near zero field. The quadratic dependence on $B$ originates from the semiclassical motion of carriers deflected by the Lorentz force. From this dependence, the mobility $\mu$ is estimated at 273 cm$^{2}$/V$\cdot$s using the Kohler's rule\cite{Kohler}, $R(B)/R(B = 0) \approx 1+(\mu B)^{2}$.

In order to take a closer look of the peak near zero field, the MR in a small field range within $\pm 1$ T is plotted in Fig. \ref{Fig3}(b), together with the data at temperatures of 3 - 10 K. As shown in the figure, the MR curves at 2 K and 3 K consist of two components; a negative MR forming a gentle hill in the range of $\pm 0.5$ T and a positive component forming a sharp dip in the range of $\pm 0.2$ T. At higher temperatures of 6 K and 10 K, only a positive MR appears. The positive MR is attributed to the WAL in the surface transport through the SS. On the other hand, the negative MR could be interpreted as a result of the weak localization (WL) in the transport through the bulk subbands; due to a small thickness of the film in the present measurement, the bulk state inside the film is quantized into 2D subbands and the WL can be expected in the transport through these quantized subbands\cite{Lu}.
The MR curves shown in Fig. 3(b) are plotted in the form of the magnetoconductance (MC) $\Delta G(B) = 1/R(B) - 1/R(B=0)$ in Fig. 4.

\begin{figure}[h]
\begin{center}
\begin{minipage}{35pc}
\begin{center}
\includegraphics[width=29pc]{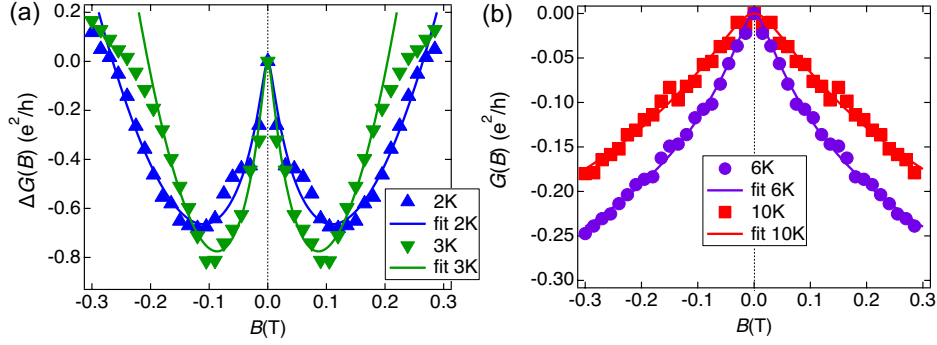}
\caption{\label{Fig4}The magnetoconductance (MC) $\Delta G(B)$ in the range of $\left| B \right| \leq$ 0.3 T. (a) Blue (green) marks represent the experimental data at 2 K (3 K). Lines of the respective colors represent the result of fitting to the HLN equation. (b) Purple (red) marks represent the experimental data at 6 K (10 K) and lines represent the result of fitting.}
\end{center}
\end{minipage} 
\end{center}
\end{figure}

According to the theory of quantum correction in the 2D diffusive transport, the MC due to the WAL or WL is given by the Hikami-Larkin-Nagaoka (HLN) equation as follows\cite{Hikami},

\begin{equation}
\label{eq1}
\Delta\sigma=\sigma(B)-\sigma(0)=-\frac{\alpha e^2}{2\pi^2 \hbar}\biggl[\ln\frac{\hbar}{4Bel^{2}_{\phi}}-\psi
\biggl(\frac{1}{2}+\frac{\hbar}{4Bel^{2}_{\phi}}\biggr)\biggr],
\end{equation}
where $e$ is the charge of an electron, $\hbar$ is the Plank's constant, $l_{\phi}$ is the phase coherent length, and $\psi(x)$ is 
the digamma function. $\alpha$ is a coefficient which should be $-$0.5 per single transport channel for the symplectic case of the WAL, and 1 for the case of the WL, respectively. 
$\Delta G(B)$ at 2 K and 3 K, represented by blue and green marks respectively in Fig. \ref{Fig4}(a), is fitted with Eq. (\ref{eq2}), expressing the sum of the contributions of the WAL and WL.
\begin{equation}
\label{eq2}
\Delta\sigma=\sigma(B)-\sigma(0)=-\frac{\alpha e^2}{2\pi^2 \hbar}\biggl[\ln\frac{\hbar}{4Bel^{2}_{\phi}}-\psi
\biggl(\frac{1}{2}+\frac{\hbar}{4Bel^{2}_{\phi}}\biggr)\biggr]-\frac{\beta e^2}{2\pi^2 \hbar}\biggl[\ln\frac{\hbar}{4Bel^{2}_{\xi}}-\psi
\biggl(\frac{1}{2}+\frac{\hbar}{4Bel^{2}_{\xi}}\biggr)\biggr],
\end{equation}
where, the first and second terms correspond to the MCs due to the WAL and WL, with the respective fitting parameters $\alpha$, $l_{\phi}$ for the WAL and $ \beta$, $l_{\xi}$ for the WL.
The parameters obtained from the fitting at 2 K (3 K) are $\alpha$ = $-$1.67 ($-$2.13), $l_{\phi}$ = 224 nm (235 nm), and $\beta$ = 276 (291), $l_{\xi}$ = 15 nm (17 nm), respectively.

\begin{figure}[h]
\begin{center}
\begin{minipage}{35pc}
\begin{center}
\includegraphics[width=16pc]{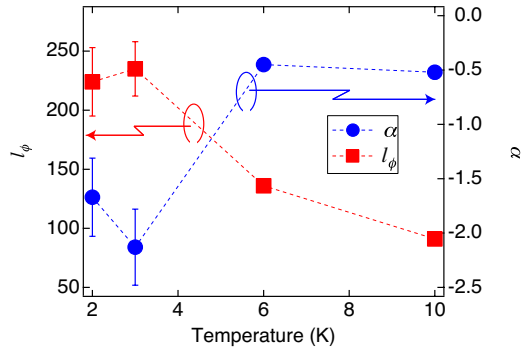}
\caption{\label{Fig5}The fitting parameters $\alpha$ and $l_{\phi}$ obtained from the fitting are plotted as a function of temperature.}
\end{center}
\end{minipage} 
\end{center}
\end{figure}

$\Delta G(B)$ at temperatures of 6 K and 10 K are plotted in Fig. \ref{Fig4}(b). Since a negative MR due to the WL is missing, 
$\Delta G(B)$ is fitted with Eq. (\ref{eq1}). The parameters obtained from the fitting at 6 K (10 K) are $\alpha$ = $-$0.45 ($-$0.52), and $l_{\phi}$ = 131 nm (86 nm), respectively. In Fig. \ref{Fig5}, the fitting parameters $\alpha$ and $l_{\phi}$ for the WAL are plotted as a function of temperature. With the increase of temperature, $\alpha$ approaches $-$0.5 and the phase coherent length $l_{\phi}$ decreases. The decrease of the absolute value of $\alpha$ with temperature suggests that the number of the transport channel through the SS decreases. 

In the surface transport through the SS of 3D TIs, which is classified into the symplectic group, the value of $\alpha$ should be $-$0.5 per single transport channel. In the (111) surface of SnTe, the total number of Dirac cones is four, three on the $\overline{\rm M}$ points which are projections of L points, and one on the $\overline{\rm \Gamma}$ point which are projections of $\Gamma$ or L points\cite{Tanaka2}. In addition, considering the surface transport through both the top and bottom surfaces of the film, the total number of the channel should be 8, resulting in the value of $\left| \alpha \right|$ = 4. However,  $\left| \alpha \right|$ around 2 was obtained from the measurement at 2K, giving the number of channels in the SS around 4 . A possible explanation for this deviation in the value of $\alpha$ would be given by the coupling between the SS and the bulk state or the coupling between the different SSs, as recently suggested for a small value of $\left| \alpha \right|$ \cite{Assaf} \cite{Steinberg}; if there is a coupling between different transport channels due to the scattering of carriers from one transport channel to another with keeping the phase coherence, they contribute to the conductance as a single phase-coherent channel.

\section{Summary}
We investigated the MR in the SnTe(111) thin film at low temperatures in order to clarify the properties of the SS of TCIs from the viewpoint of electrical transport.
We observed that positive and negative components coexist in the MR curve near zero field at lowest temperatures of 2 - 3 K while only the positive MR appears at elevated temperatures of 6 - 10 K. The positive MR is attributed to the WAL originating from the surface transport in the SS, while the negative MR to the weak localization (WL) effect originating from the transport in the quantized 2D subbands of the bulk state. We deduced the prefactor $\alpha$ and the phase coherence length $l_{\phi}$ by fitting the observed MC due to the WAL with the HLN equation. The value of $\left| \alpha \right|$ was much smaller than expected from the nominal total number of the transport channels of the SS in the SnTe (111) surface, which suggests  the coupling between the SS and bulk state.

\ack{}
The authors would like to thank T. Koyano in the Cryogenics Division of Research Facility Center for Science and Technology, University of Tsukuba for PPMS measurements, and T. Suemasu and K. Ito for AFM measurements. This work is partially supported by Grant-in-Aid for Scientific Research (KAKENHI) from the Japan Society for the Promotion of Science, CASIO Science Promotion Foundation, and Sumitomo Electric Industries Group CSR Foundation.

\end{document}